\pdfoutput=1
\documentclass[11pt,twoside,a4paper,cmspaper,final,collab]{cms-tdr}
\def\svnVersion{2dcfa6c}\def\svnDate{2023/10/03}\def\cmsCernNoTag{CERN-EP-2022-131}\def\cmsCernDate{\today}\def\cmsMessage{Published in Physics Letters B as \href{http://dx.doi.org/10.1016/j.physletb.2023.137783}{\doi{10.1016/j.physletb.2023.137783}.}}
\begin{document}\cmsNoteHeader{HIG-21-015}

\newcommand{\pp}{\ensuremath{\Pp\Pp}\xspace}
\newcommand{\sqrts}{\ensuremath{\sqrt{s}=13\TeV}\xspace}
\newcommand{\intlumi}{\ensuremath{138\fbinv}\xspace}
\newcommand{\Hee}{\ensuremath{\PH\to\Pep\Pem}\xspace}
\newcommand{\Heeg}{\ensuremath{\PH\to\Pep\Pem\PGg}\xspace}
\newcommand{\Hmumu}{\ensuremath{\PH\to\Pgmp\Pgmm}\xspace}
\newcommand{\BHee}{\ensuremath{\mathcal{B}(\Hee)}\xspace}
\newcommand{\mee}{\ensuremath{m_{\Pe\Pe}}\xspace}
\newcommand{\ee}{\ensuremath{\Pep\Pem}\xspace}
\newcommand{\mH}{\ensuremath{m_\PH}\xspace}
\newcommand{\Hgg}{\ensuremath{\PH\to\PGg\PGg}\xspace}
\newcommand{\Zee}{\ensuremath{\PZ\to\Pep\Pem}\xspace}
\newcommand{\Zmumu}{\ensuremath{\PZ\to\PGmp\PGmm}\xspace}
\newcommand{\VH}{\ensuremath{\PV\PH}\xspace}
\newcommand{\ttH}{\ensuremath{\ttbar\PH}\xspace}
\newcommand{\seff}{\ensuremath{\sigma_{\text{eff}}}\xspace}
\newcommand{\ggH}{\ensuremath{\Pg\Pg\PH}\xspace}
\newcommand{\muR}{\ensuremath{\mu_{\mathrm{R}}}\xspace}
\newcommand{\muF}{\ensuremath{\mu_{\mathrm{F}}}\xspace}

\ifthenelse{\boolean{cms@external}}{\providecommand{\cmsTable}[1]{#1}}{\providecommand{\cmsTable}[1]{\resizebox{\textwidth}{!}{#1}}}
\newlength\cmsTabSkip\setlength{\cmsTabSkip}{1ex}
\newlength\cmsFigWidth
\ifthenelse{\boolean{cms@external}}
{\setlength{\cmsFigWidth}{0.49\textwidth}}{\setlength{\cmsFigWidth}{0.75\textwidth}}
\cmsNoteHeader{HIG-21-015}
\title{Search for the Higgs boson decay to a pair of electrons in proton-proton collisions at \texorpdfstring{\sqrts}{sqrt(s) = 13 TeV}}

\date{\today}

\abstract{
  A search is presented for the Higgs boson decay to a pair of electrons (\ee) in proton-proton collisions at \sqrts.
  The data set was collected with the CMS experiment at the LHC between 2016 and 2018, corresponding to an integrated luminosity of \intlumi.
  The analysis uses event categories targeting Higgs boson production via gluon fusion and vector boson fusion.
  The observed upper limit on the Higgs boson branching fraction to an electron pair is $3.0\times10^{-4}$ ($3.0\times10^{-4}$ expected) at the 95\% confidence level, which is the most stringent limit on this branching fraction to date.
}

\hypersetup{
  pdfauthor={CMS Collaboration},
  pdftitle={Search for the Higgs boson decay to a pair of electrons in proton-proton collisions at sqrt(s) = 13 TeV},%
  pdfsubject={CMS},
  pdfkeywords={CMS, Higgs, electrons, BSM}
}

\maketitle

\section{Introduction}
\label{sec:intro}

Since the discovery of the Higgs boson by the ATLAS and CMS Collaborations in 2012~\cite{Aad:2012tfa,Chatrchyan:2012xdj,Chatrchyan:2013lba}, various measurements of its interactions with standard model (SM) particles have been performed.
In particular, the interactions of the Higgs boson with the third-generation fermions have been observed, with coupling strengths found to be in agreement with the expectations of the SM~\cite{CMSttH,ATLASttH,CMSHbb,ATLASHbb,CMSHtautau,ATLASHtautau}.
More recently, the CMS Collaboration has also presented first evidence for the decay of the Higgs boson to the second-generation fermions, in a search for \Hmumu decays~\cite{CMSHMuMu}.
Such measurements of Yukawa couplings to light fermions are experimentally challenging at the LHC, given that the SM predicts the coupling strengths to be proportional to the fermion mass~\cite{SM1,SM2,SM3}.
Consequently, couplings of the Higgs boson to the first-generation fermions have yet to be confirmed experimentally.

For the Higgs boson decay to an electron pair, \Hee, the SM predicted branching fraction of $\BHee \approx 5\times10^{-9}$~\cite{YR3} is extremely small and inaccessible with the current LHC data set.
Higher-order corrections included in the branching fraction prediction introduce diagrams which do not contain the Higgs boson Yukawa coupling to the electron. 
These contributions are expected to dominate in the SM, but are still much smaller than present sensitivity. 
Therefore, a search for this decay channel currently provides the only direct probe of the Higgs boson Yukawa coupling to the electron, which is enhanced in several scenarios beyond the SM.
The simplest of these are models postulating two Higgs doublets~\cite{BSM_2HDM}; other extensions include the addition of higher order operators to the SM Lagrangian, including dimension 10 operators that could modify the coupling by a factor of ${\approx}10$~\cite{BSM_vecferm,dim_ten_operators}. These enhancements are, however, still smaller than present sensitivity could detect.

This Letter describes a search for \Hee decays in proton-proton (\pp) collisions at \sqrts, using data recorded with the CMS detector in 2016--2018, corresponding to an integrated luminosity of \intlumi.
The most recent search from the CMS Collaboration for this decay was performed using \pp collision data collected at $\sqrt{s}=8\TeV$ with an integrated luminosity of 19.7\fbinv~\cite{CMSHeeRun1}.
A 95\% confidence level (\CL) upper limit on \BHee of $1.9\times10^{-3}$ was determined.
The most sensitive search for \Hee decays from the ATLAS Collaboration was performed using \pp data with an integrated luminosity of 139\fbinv, collected at $\sqrt{s}=13\TeV$.
A 95\% CL upper limit on \BHee was set at $3.6\times10^{-4}$~\cite{ATLASHeeRun2}.

The analysis presented in this Letter primarily targets Higgs boson production by gluon fusion (\ggH) and vector boson fusion (VBF).
Rarer Higgs boson production processes, including production in association with top quarks (\ttH) and a vector boson (\VH), are also considered, although no dedicated selection is applied to target them.
The final states of interest comprise a pair of two prompt, oppositely charged electrons, potentially produced in association with hadronic jets.
The sensitivity to each production mode is increased by dividing events into analysis categories, giving an improved signal to background ratio (S/B).
Each analysis category is defined by a selection on a dedicated multivariate (MVA) discriminant.
The MVA-based classifiers are trained to separate Higgs boson signal events from the dominant Drell--Yan (DY) background and smaller background contributions from \ttbar decays.
A simultaneous fit to the dielectron invariant mass distribution (\mee) in each category is then used to extract an upper limit on \BHee.

\section{The CMS detector}
\label{sec:detector}

The central feature of the CMS apparatus is a superconducting solenoid of 6\unit{m} internal diameter, providing a magnetic field of 3.8\unit{T}.
Within the solenoid volume are a silicon pixel and strip tracker, a lead tungstate crystal electromagnetic calorimeter (ECAL), and a brass and scintillator hadron calorimeter (HCAL), each composed of a barrel and two endcap sections.
The pseudorapidity ($\eta$) coverage provided by the barrel and endcap detectors is extended by the forward hadron calorimeter, which uses steel as an absorber and quartz fibres as the sensitive material.
Muons are detected in gas-ionization chambers embedded in the steel flux-return yoke outside the solenoid.

Events of interest are selected using a two-tiered trigger system~\cite{CMS_L1T,CMS_HLT}. The first level (L1), composed of custom hardware processors, uses information from the calorimeters and muon detectors to select events at a rate of around 100\unit{kHz} within a fixed latency of about 4\mus.
The second level, known as the high-level trigger, consists of a farm of processors running a version of the full event reconstruction software optimized for fast processing, and reduces the event rate to around 1\unit{kHz} before data storage.

The global event reconstruction (also called particle-flow (PF) event reconstruction~\cite{ParticleFlow}) aims to reconstruct individual particles (photons, charged and neutral hadrons, muons, and electrons) by optimally combining information from the various elements of the CMS detector.
In the case of electrons, the energy is determined from a combination of the electron momentum at the primary interaction vertex as determined by the tracker, the energy of the corresponding ECAL cluster, and the energy sum of all bremsstrahlung photons spatially compatible with originating from the charged electron track.
The momentum resolution for electrons with transverse momentum $\pt \approx 45\GeV$ from \Zee decays ranges from 1.6--5.0\%~\cite{CMS-DP-2020-021,CMS_egamma_performance}.
The candidate vertex with the largest value of summed physics-object $\pt^2$ is taken to be the primary \pp interaction vertex.
The physics objects used for this determination are the jets, clustered using the infrared- and collinear-safe anti-\kt algorithm~\cite{anti-kt,fastJet} with the tracks assigned to candidate vertices as inputs, and the associated missing transverse momentum, taken as the negative vector \pt sum of those jets.

For charged hadrons, the energy is determined from a combination of their momentum measured in the tracker and the matching ECAL and HCAL energy deposits, corrected for the response function of the calorimeters to hadronic showers.
The energy of neutral hadrons is obtained from the corresponding corrected ECAL and HCAL energies.
Hadronic jets are built from PF candidates using the anti-\kt algorithm~\cite{anti-kt} with a distance parameter of 0.4.
The momentum of a jet is determined as the vectorial sum of all particle momenta in the jet, and is found from simulation to be, on average, within 5--10\% of the true momentum over the whole \pt spectrum and detector acceptance.
Charged hadrons originating from additional \pp interactions are removed from the analysis.

A more detailed description of the CMS detector, together with a definition of the coordinate system used and the relevant kinematic variables, can be found in Ref.~\cite{CMSdetector}.

\section{Analysis strategy}
\label{sec:strategy}

The analysis strategy is similar to that used in the SM \Hgg analysis~\cite{HIG_19_015}.
In this case, however, events are selected using a dielectron trigger with asymmetric electron \pt thresholds of 23 and 12\GeV, and loose isolation requirements.
Offline, all events entering the analysis are required to pass a loose selection placed on the electron kinematics and an electron identification (ID) criterion.
Simulated signal and background events are then used to train MVA-based classifiers designed to reduce the number of background events entering the selection.
This is performed independently for the \ggH and VBF channels, with classifiers trained on simulated samples corresponding to the three data-taking years, combined in accordance with their integrated luminosities.

In both \ggH and VBF channels, the dominant background consists of DY events, with smaller contributions from \ttbar processes with dilepton or lepton + jets final states.
In the VBF channel, electroweak \PZ boson production in association with two jets (EW \Zee) is also considered.
These events closely mimic the VBF signal process and are included as a background during MVA training.
Although the cross section for VBF production is roughly an order of magnitude smaller than for \ggH production, the VBF final state provides a distinct signature in the detector, in which the Higgs boson is produced in association with two forward jets with a large $\eta$ difference and high dijet invariant mass.
These features are exploited by the VBF classifier to enhance the expected S/B ratio, allowing the sensitivity of the VBF categories to become comparable to those targeting \ggH.

The output scores of the classifiers are used to define analysis categories.
The number and location of category boundaries is chosen to optimize the expected significance of all resulting categories combined.
The categories are non-overlapping by construction, with events first being considered for the VBF categories, as these have both a higher S/B and a distinctive topology.
If an event fails the VBF requirements, it is then considered for the \ggH analysis categories.

Once the analysis categories are defined, a maximum likelihood fit is performed to the \mee distribution in each category simultaneously.
The smoothly-falling background spectrum, consisting mostly of DY events, is modelled directly from data using the discrete profiling method~\cite{envelope}.
The signal model is derived from simulated Higgs boson events independently for each production mode and analysis category.
The fit is used to extract 95\% \CL upper limits on \BHee.

\section{Data and simulated events}
\label{sec:samples}

This analysis uses \pp collision data corresponding to an integrated luminosity of \intlumi, of which 36.3, 41.5, and 59.8\fbinv were collected during 2016, 2017, and 2018, respectively.

A Monte Carlo (MC) simulated sample is generated for each of the \ggH, VBF, \VH, and \ttH signal processes using \MGvATNLO{} v2.6.5~\cite{madgraph5} at next-to-leading order (NLO) accuracy in perturbative quantum chromodynamics (QCD).
For the \ggH process, up to three additional partons are considered in the matrix element calculation; for VBF production, up to one additional parton is considered, excluding the quark-initiated jets produced in the leading order process.
In addition, events from \ggH production are weighted to match the predictions of the \textsc{nnlops} program~\cite{Hamilton:2013fea}, as a function of the Higgs boson transverse momentum and the number of jets in the event.
Simulated DY and electroweak \Zee background processes are also generated at NLO using \MGvATNLO{}, while \ttbar events are produced using the \POWHEG{}~v2.0~\cite{powheg1,powheg2,powheg3,powheg_ttbar} NLO generator.
All parton-level simulated samples are interfaced with \PYTHIA{}~v8.230~\cite{pythia8} for parton showering and hadronization, using the CP5 underlying event tune~\cite{CP5Tune}.
Parton distribution functions (PDFs) are taken from the NNPDF~3.1~\cite{NNPDF31} set for the simulation of all years.
The \ggH, VBF, \VH, and \ttH production cross sections used to normalize the signal samples are taken from the LHC Higgs Working Group recommendations~\cite{YR4}, at the highest available order.
Additional \pp interactions occurring in both the same and adjacent bunch crossings (pileup) are simulated as a set of minimum bias interactions that are mixed with the hard scattering event.
In the analysis of each data-taking year, the simulated events are weighted based on the number of pileup events to match the distribution measured in data.
Finally, the response of the CMS detector is simulated using the \GEANTfour{} package~\cite{geant4}.

\section{Event selection}
\label{sec:selection}

A loose preselection on the events is applied to ensure they are consistent with a Higgs boson decaying to two electrons.
The two electrons with the highest and next-highest \pt are referred to as the leading and subleading electrons, respectively, and define the dielectron system.
Each event must contain at least two oppositely charged electrons fulfilling the following requirements:

\begin{itemize}
  \item $\pt > 35$ (25)\GeV for the leading (subleading) electron;
  \item a tight working point on an electron ID, designed to have a 90\% signal selection efficiency.
        The ID combines descriptions of the shape and isolation of energy deposits in the ECAL, the quality of associated tracks, and the compatibility of measurements from the tracker and ECAL, in an MVA-based discriminator~\cite{CMS_egamma_performance};
  \item pseudorapidity within the ECAL acceptance ($\abs{\eta}<2.5$), and not in the barrel-endcap transition region ($1.44<\abs{\eta}<1.57$); and
  \item $110 < \mee < 150\GeV$, chosen to limit contributions from DY events, and to ensure the \mee sideband regions have sufficient events to constrain the background expectation in the signal region.
\end{itemize}

All jets entering the analysis are required to have $\pt>25\GeV$ and $\abs{\eta} < 4.7$.
Jets are required to pass a tight pileup identification criterion~\cite{PUJID} that uses the topology of the jet shape, the number of charged and neutral jet constituents, and information on any associated tracks, to reject jets resulting from pileup.
A tight requirement is also placed on an additional jet ID, which rejects spurious jets resulting from detector noise~\cite{PUJID}.
Finally, a selection is applied to suppress the observed noise in the ECAL endcaps for low-\pt jets in the 2017 data; this selection vetoes jets with $\pt<50\GeV$ within the region $2.7<\abs{\eta}<3.1$.

Scale factors are applied to simulated samples to correct for differences between simulation and data.
These include corrections for differences in the electron reconstruction and identification efficiencies, as well as in the trigger efficiency.

\section{Event categorization}
\label{sec:categorisation}

The categorization targeting VBF events is based on a boosted decision tree (BDT) trained to discriminate between signal and background events, referred to as the VBF BDT.
Prior to training the BDT, an additional selection is applied on top of the loose preselection in order to target VBF events.
This is referred to as the VBF preselection and comprises the following:

\begin{itemize}
  \item $\pt > 40$ (25)\GeV for the leading (subleading) jet;
  \item dijet invariant mass $>350\GeV$.
\end{itemize}

All simulated VBF events passing the VBF preselection are considered as signal when training the BDT.
The largest background process in the VBF phase space consists of DY events, with a smaller but significant contribution from \ttbar production.
Although small in yield, the electroweak \PZ boson production process is also considered, since such events typically have signal-like characteristics and thus are assigned high VBF BDT scores.

\begin{figure*}[tb!]
  \centering
  \includegraphics[width=0.49\textwidth]{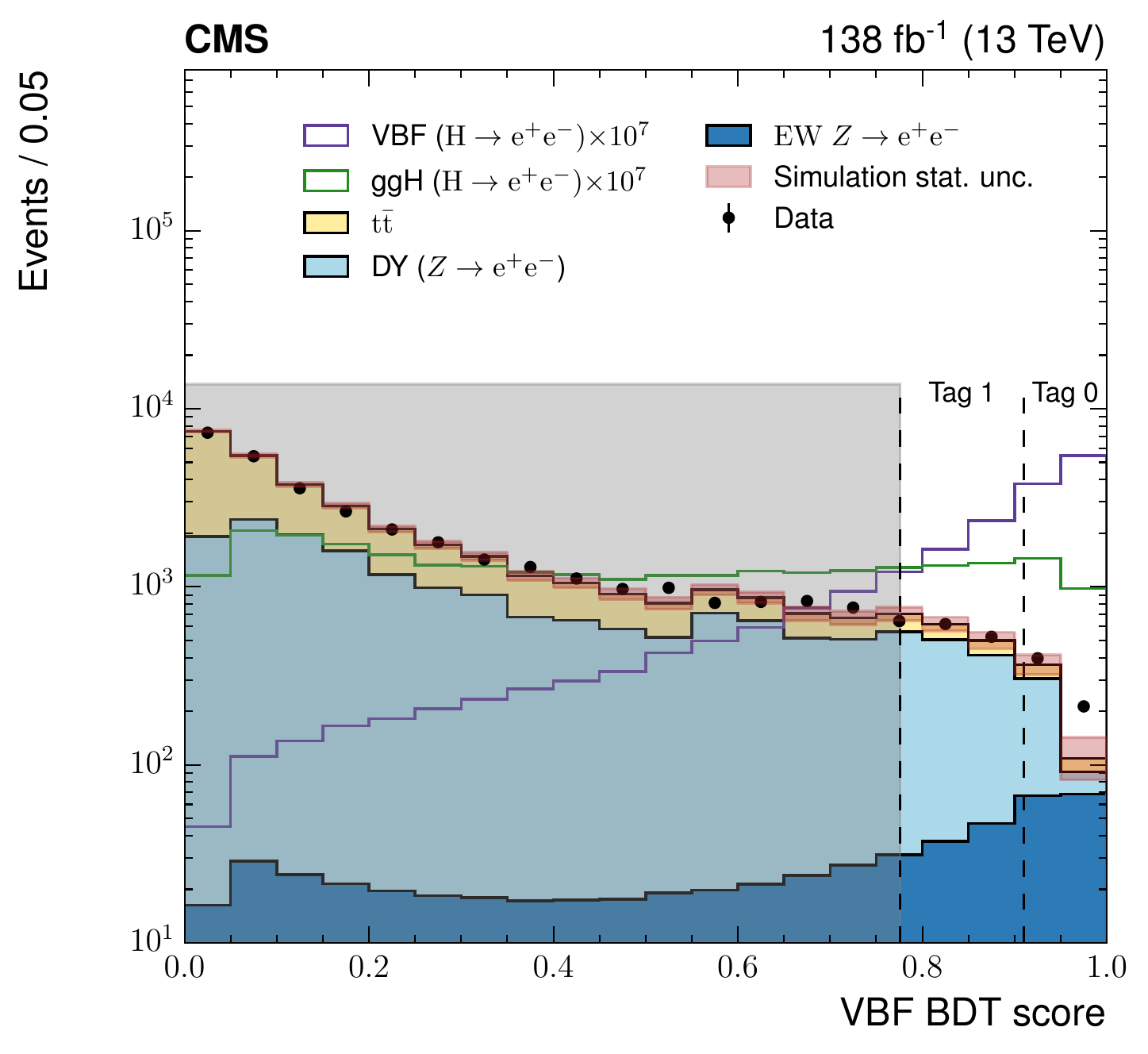}\hfill%
  \includegraphics[width=0.49\textwidth]{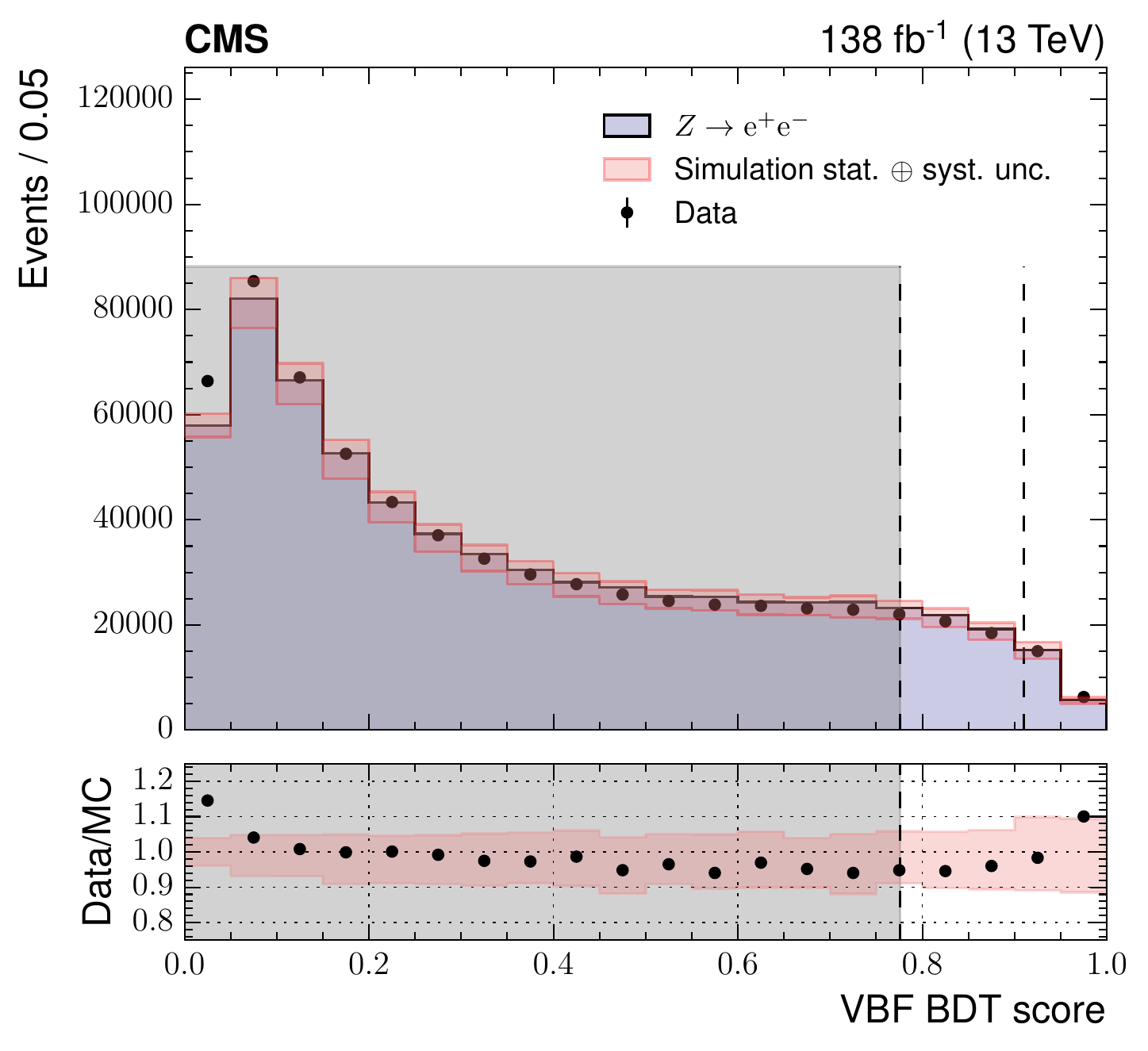}\hfill%
  \caption{
    Distribution of the output score of the VBF BDT in all simulated background and signal events, and data (left), passing VBF preselection.
    The \ggH and VBF signals are scaled for better visibility.
    Category boundaries targeting VBF production are denoted with dashed lines.
    The shaded region defines events which are not selected to enter VBF analysis categories, but may populate those targeting \ggH.
    A small excess in data with respect to the MC background prediction is observed in the most signal-like bin; however, these events are not found to peak around the signal region of the \mee distribution.
    The right plot shows the distribution of the output score of the VBF BDT in a control region around the \PZ boson mass.
    The simulated samples include both DY and EW \Zee events.
    The combined impact of the statistical and systematic uncertainties in simulation is shown by the red shaded band, where the systematic component includes uncertainties in the jet energy scale and resolution corrections, alongside the electron energy scale corrections.
    Uncertainties in the efficiency of electron identification, reconstruction, and trigger selection are also included, as well as the uncertainty in the integrated luminosity, presented in Section~\ref{sec:systematics}.
    Good agreement is observed between the DY simulation (filled histogram) and data (black markers), within the phase space in which the analysis categories are constructed.
  }
  \label{fig:VBF_score_and_valid}
\end{figure*}

Many observables characterising the VBF event topology are leveraged in the BDT in order to significantly increase the S/B in each analysis category, including information related to the kinematics of the individual electrons, the dielectron system, the three leading jets individually, the dijet system, and the dielectron plus dijet system.
Inputs with good discriminating power include the invariant mass of the dijet object, the difference in $\eta$ between the two leading jets, and a centrality variable~\cite{Zeppenfeld} defined as $\exp[-4(\abs{\eta_{\Pe\Pe} - \frac{1}{2}(\eta_{j_{1}}+\eta_{j_{2}})} /\abs{\eta_{j_{1}}-\eta_{j_{2}}})^2]$, where $\eta_{\Pe\Pe}$ is the dielectron pseudorapidity, and $j_{1(2)}$ labels the leading (subleading) jet.
Other jet variables such as a quark-gluon identification score~\cite{JetsInRun2} are also included.
The score is based on quantities such as the \pt and multiplicity of PF candidates reconstructed within a jet in a likelihood discriminant to separate gluon and quark-initiated jets.
Finally, it was verified that information related to the dielectron system was not sufficient for the Higgs boson mass (\mH) to be inferred by the classifier.
The output score for the VBF BDT is shown for both simulation and data in Fig.~\ref{fig:VBF_score_and_valid} (left).

In this analysis, the background model is taken directly from data, whereas the signal modelling uses simulated samples.
Therefore, it is necessary to ensure there is reasonable agreement between data and simulation for input observables to the VBF classifier as well as the output score.
This is checked using \Zee events from a control region, defined to be non-overlapping with the analysis category phase space.
The VBF preselection is also applied, with the exception of the \mee requirements which are shifted to approximately centre on the \PZ boson mass ($80<\mee<100\GeV$).
Drell--Yan events are chosen for this validation since the final state mimics that of the \Hee process and is relatively free from contaminating backgrounds.
The distribution of the classifier output score is shown in Fig.~\ref{fig:VBF_score_and_valid} (right).
Agreement between data and simulation is observed to be within the uncertainty given by the combined systematic and statistical variations, for the range of output scores in which the analysis categories are constructed.

Analysis categories targeting VBF events are defined using the output score of the BDT.
The analysis defines only two VBF categories, since the sensitivity to VBF events is observed to saturate at numbers larger than two.
The expected signal and background yields for both VBF categories are shown in Table~\ref{tab:s_b}.

\begin{table*}[htbp!]
  \topcaption{
    The total expected number of signal events for $\mH=125.38\GeV$ in analysis categories targeting \ggH and VBF events, for an integrated luminosity of \intlumi.
    The fractional contribution from each production mode to each category is also shown.
    The \seff, defined as the smallest interval containing 68.3\% of the \mee distribution, is listed for each analysis category.
    The final column shows the expected ratio of signal to background, where S and B are the numbers of expected signal and background events in a $\pm1\seff$ window centred on \mH. The expected background event yields are extrapolated from data sidebands.
  }
  \label{tab:s_b}
  \centering
  \cmsTable{
    \begin{tabular}{rcccccccc}
      \multirow{3}{*}{Analysis categories} & \multicolumn{6}{c}{SM 125.38\GeV expected \Hee signal yields} & \multirow{3}{*}{Bkg. per \GeVns}              & \multirow{3}{*}{S/B}                                                                               \\
                                           & \multirow{2}{*}{\begin{tabular}[c]{@{}c@{}}Yield\\($\times 10^{-4}$)\end{tabular}}                & \multicolumn{4}{c}{Production mode fractions} & \multirow{2}{*}{\begin{tabular}[c]{@{}c@{}}\seff\\(\GeVns)\end{tabular}} &                                                   \\
                                           &                                                               & \ggH                                          & VBF                                            & \VH   & \ttH  &      &       &                    \\ \hline
      \ggH Tag 0                           & 4.7                                                           & 81.0\%                                        & 12.1\%                                         & 5.0\% & 1.9\% & 1.65 & 250   & $7.7\times10^{-7}$ \\
      [\cmsTabSkip]
      \ggH Tag 1                           & 18                                                            & 86.5\%                                        & 8.4\%                                          & 4.2\% & 0.9\% & 1.84 & 2340  & $2.8\times10^{-7}$ \\
      [\cmsTabSkip]
      \ggH Tag 2                           & 35                                                            & 91.7\%                                        & 4.8\%                                          & 3.1\% & 0.5\% & 2.02 & 8760  & $1.4\times10^{-7}$ \\
      [\cmsTabSkip]
      \ggH Tag 3                           & 61                                                            & 92.6\%                                        & 3.5\%                                          & 3.2\% & 0.7\% & 2.52 & 30500 & $5.4\times10^{-8}$ \\
      [\cmsTabSkip]
      VBF Tag 0                            & 2.2                                                           & 19.8\%                                        & 80.0\%                                         & 0.1\% & 0.1\% & 2.04 & 25.9  & $2.8\times10^{-6}$ \\
      [\cmsTabSkip]
      VBF Tag 1                            & 1.7                                                           & 40.3\%                                        & 58.5\%                                         & 0.8\% & 0.4\% & 2.26 & 81.7  & 6.3$\times10^{-7}$ \\
      [\cmsTabSkip]
    \end{tabular}
  }
\end{table*}

\begin{figure*}[htb!]
  \centering
  \includegraphics[width=0.49\textwidth]{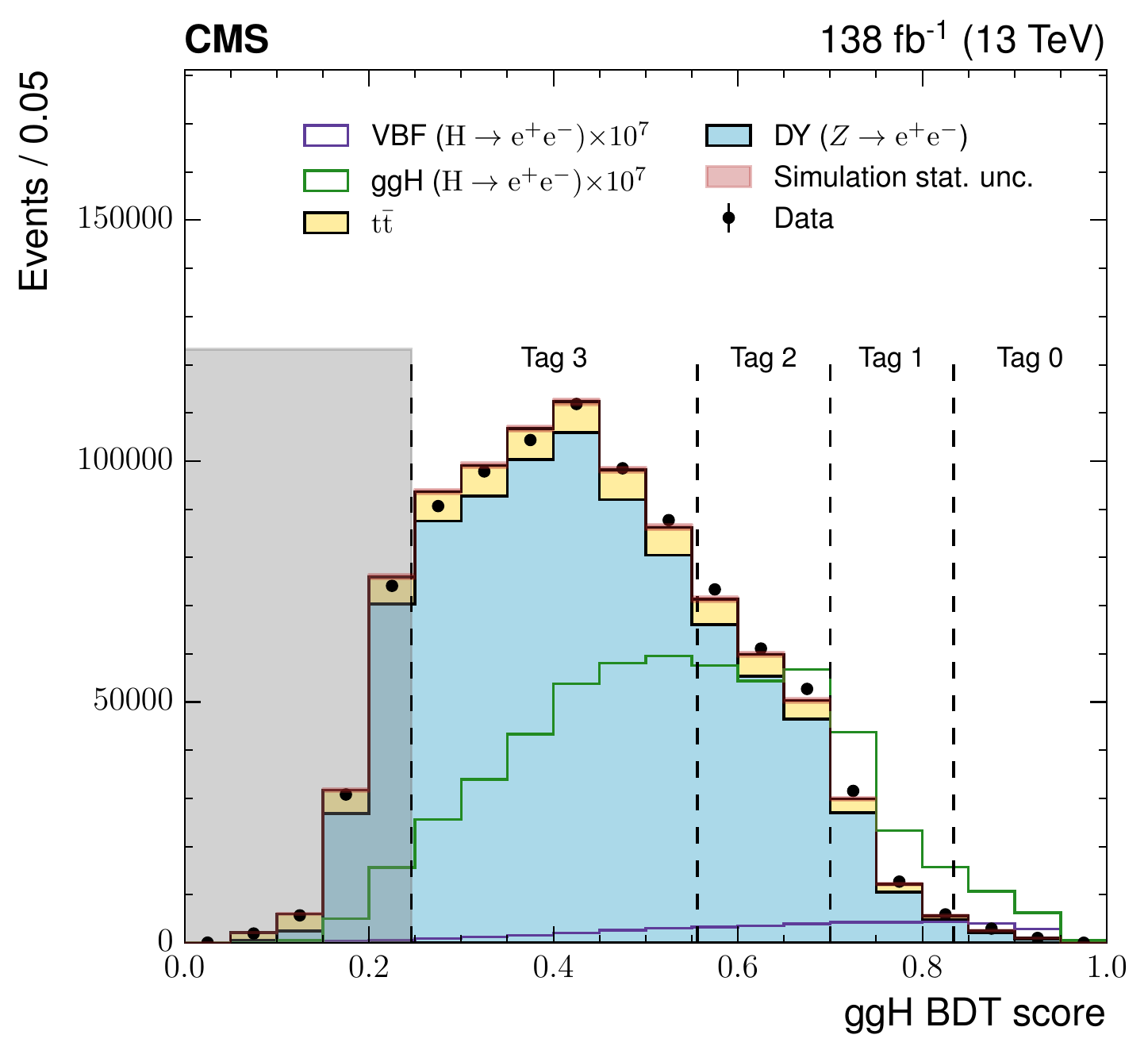}\hfill
  \includegraphics[width=0.49\textwidth]{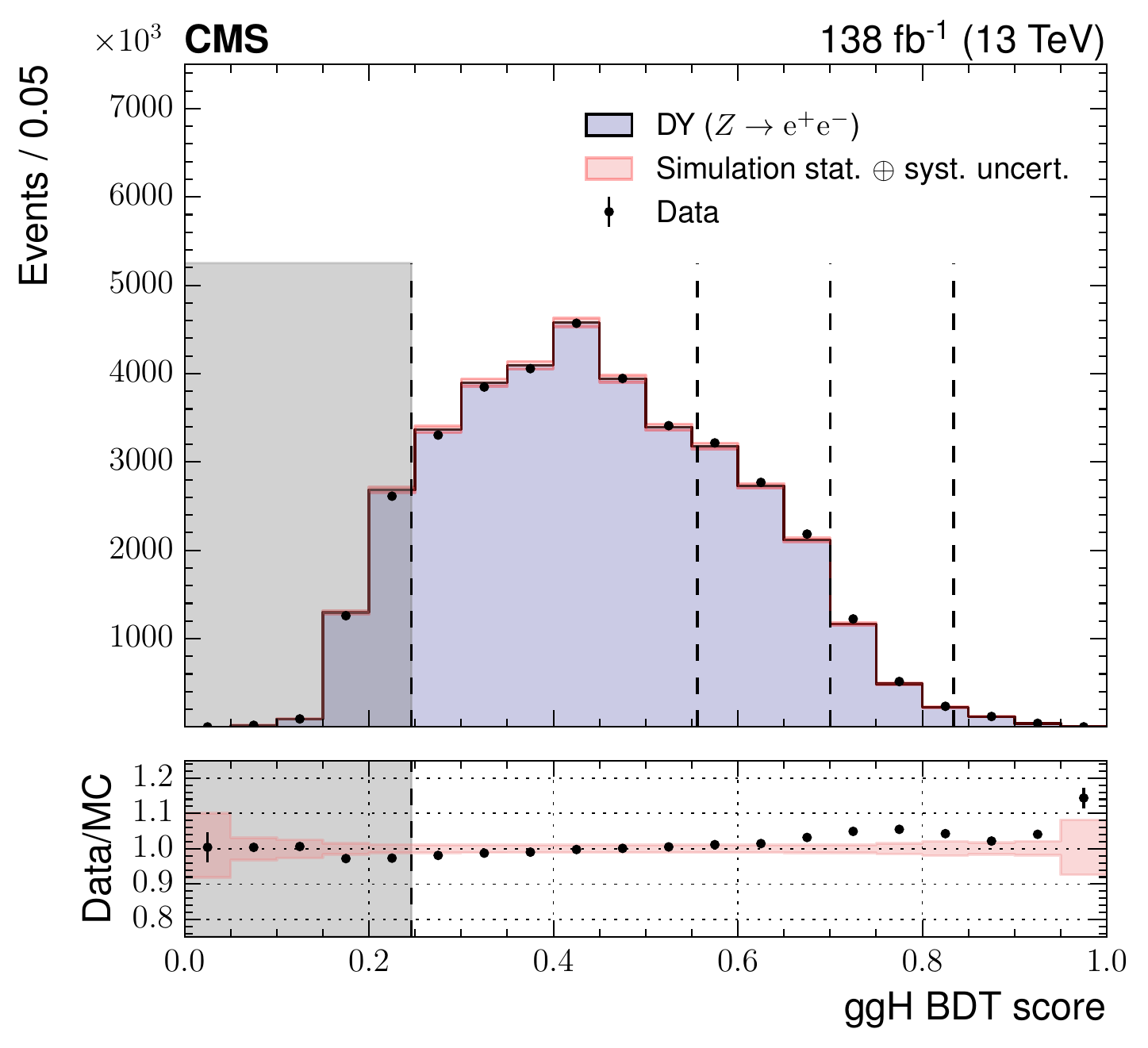}\hfill
  \caption{
    Distribution of the output score of the \ggH BDT in simulated background and signal events, and data (left).
    The \ggH and VBF signals are scaled such that they are visible.
    Category boundaries targeting \ggH Higgs boson production are denoted with dashed lines.
    Events with scores in the grey shaded region are discarded from the analysis.
    The right plot shows the distribution of the output score of the \ggH BDT in a control region around the \PZ boson mass.
    Agreement is compared between the DY simulation (filled histogram) and data (black points).
    The combined impact of the statistical and systematic uncertainties in simulation is shown by the red shaded band, where the sources contributing to the systematic component are identical to those included in Fig.~\ref{fig:VBF_score_and_valid}.
    Residual differences between data and simulation are smaller than the theory uncertainties in the \ggH cross section and event kinematics, which are included in the final maximum likelihood fit.
  }
  \label{fig:ggH_score_and_valid}
\end{figure*}

The categorization for \ggH events is also based on a BDT trained to discriminate between simulated signal and background events, referred to as the \ggH BDT.
All simulated signal and background events passing the basic preselection are considered when training.
The largest background contribution in this channel consists of DY events, the kinematics of which are typically similar to the \ggH signal.
Hence, the classification task is nontrivial and many features are leveraged to improve the separation power of the BDT.
These features include kinematic properties of the individual electrons, the dielectron system, and up to two jets, if present.
Although these inputs are broadly similar between simulated \ggH signal and background events, quantities such as the dielectron \pt provide reasonable discriminating power, where the average dielectron \pt for \ggH events is higher than in background.
Similarly to the VBF BDT, it is checked that information related to the dielectron system is not sufficient for \mH to be inferred by the classifier.
The output score for the \ggH BDT is shown for both simulation and data in Fig.~\ref{fig:ggH_score_and_valid} (left).

The validation procedure for the \ggH  BDT is identical to that for the VBF BDT, described above, and is performed in the \PZ boson mass control region designed to be non-overlapping with the ggH analysis categories.
The basic analysis preselection is also applied, with the exception of the \mee requirements.
The distribution of the classifier output score in this region is shown in Fig.~\ref{fig:ggH_score_and_valid} (right).
The agreement between data and simulation is typically within the uncertainty permitted by the combined systematic and statistical variations; any residual differences are smaller than the theoretical uncertainty in the \ggH production cross section, presented in Section~\ref{sec:systematics}.

Four analysis categories targeting \ggH events are defined using the output score of the \ggH BDT.
It is checked that going beyond four categories offers no improvement in sensitivity.
The expected signal and background yields in each \ggH category are shown in Table~\ref{tab:s_b}.

\section{Systematic uncertainties}
\label{sec:systematics}

In this analysis, the systematic uncertainty associated with the background estimation from data is handled using the discrete profiling method, as described in Section~\ref{sec:results}.
Systematic uncertainties that affect the signal model are implemented in one of the following two ways.
Uncertainties that modify the shape of the \mee distribution are incorporated into the signal model as nuisance parameters.
These are typically experimental uncertainties related to the energy measurement of the individual electrons.
Conversely, if the shape of the \mee distribution is unaffected, the uncertainty is treated as a log-normal constrained variation in the event yield.

\subsection{Theoretical uncertainties}
The sources of theoretical uncertainty considered in this analysis are listed below.
The effects of theoretical uncertainties are taken to be correlated across years.

\begin{itemize}
  \item \textit{Renormalization and factorization scale uncertainty}:
        the uncertainty arising from variations of the renormalization (\muR) and factorization (\muF) scales used when computing the expected SM cross section and event kinematics.
        These account for the missing higher-order terms in perturbative calculations.
        The recommendations provided by the Higgs Cross Section Working Group quoted in Ref.~\cite{YR4} are followed.
        The uncertainties in the signal acceptance due to the \muR and \muF scales are estimated using three sources: varying the \muR scale by a factor of 2 and 0.5, varying the \muF scale by a factor of 2 and 0.5, and varying both in the same direction simultaneously.
        The impacts of the signal acceptance uncertainties are evaluated keeping the overall normalization of each signal process constant, and are at largest (for the \ggH production mode) 10\%.
  \item \textit{PDF uncertainties}:
        these account for the uncertainty due to imperfect knowledge of the composition of the proton, that affects which partons are most likely to initiate high energy events.
        The normalization uncertainties are computed following the PDF4LHC prescription~\cite{PDF4LHC,YR3}.
        The impact on the normalization ranges between 1.9 and 3\%.
  \item \textit{Uncertainty in the strong force coupling constant}:
        the uncertainty in the value of the strong force coupling constant $\alpS$ is included in the treatment of the PDF uncertainties, following the PDF4LHC~\cite{PDF4LHC,YR3} prescription.
        The impact on the normalization is largest for \ggH production, with a value of 2.6\%.
  \item \textit{Underlying event and parton shower uncertainties}:
        these uncertainties are obtained using dedicated simulated samples which vary the \PYTHIA{} tune from that used in the nominal simulation samples, and vary the renormalization scale for QCD emissions in both initial-state and final-state radiation by a factor of 2 and 0.5.
        These uncertainties are treated as migrations of events from a given production mode into and out of the VBF and ggH analysis categories.
        The largest effect comes from the parton shower uncertainty in VBF events, which can change the signal yield in the VBF analysis categories by up to 5\%.
\end{itemize}

The total impact on the \BHee measurement from theoretical systematic uncertainties is ${}^{+0.2}_{-0.1} \times 10^{-4}$, with dominant contributions from uncertainties affecting the \ggH production cross section.

\subsection{Experimental uncertainties}
The uncertainties that affect the shape of the signal \mee distribution are listed below.

\begin{itemize}
  \item \textit{Electron energy scale}:
        the uncertainty associated with the corrections applied to the electron energy scale in simulated events~\cite{CMS_egamma_performance}, derived using a sample of \Zee tag-and-probe (T\&P) events~\cite{TagAndProbe}.
        Four nuisance parameters are defined for the possible combinations of low $R_{9}$, high $R_{9}$, electrons in the endcap, and electrons in the barrel region of the ECAL, where $R_{9}$ is defined as the energy sum of the $3 \times 3$ crystals centred on the most energetic crystal in the candidate electromagnetic cluster divided by the sum of the crystal energies assigned to that electromagnetic cluster.
  \item \textit{Electron energy scale nonlinearity}:
        a further source of uncertainty to cover possible differences between the linearity of the electron energy scale between data and simulation, estimated on a sample of \Zee T\&P events.
        An uncertainty of 0.1\% is assigned for electrons with $\pt<80\GeV$, and 0.2\% for those above.
\end{itemize}

The uncertainties that only modify the event yield include:
\begin{itemize}
  \item \textit{Integrated luminosity}:
        the integrated luminosities for the 2016, 2017, and 2018 data-taking years have individual uncertainties of 1.2, 2.3, and 2.5\%, respectively~\cite{CMSlumi2016,CMSlumi2017,CMSlumi2018}.
        The total uncertainty for the 2016--2018 period is 1.6\%.
        These are partially correlated across the different data sets to account for common sources of uncertainty in the luminosity measurement schemes.
  \item \textit{Electron identification and reconstruction}:
        uncertainties in the scale factors derived to correct for differences in simulation and data for the electron identification and reconstruction efficiency.
        For both sources, the size of the uncertainty is approximately 1\% in each category.
  \item \textit{Jet energy scale and resolution corrections}:
        the energy scale of jets is measured using the \pt balance of jets with \PZ bosons and photons in \Zee, \Zmumu and $\PGg$+jets events, as well as the \pt balance between jets in dijet and multijet events~\cite{JetsInRun2}.
        The uncertainty in the jet energy scale is a few percent and depends on \pt and $\eta$.
        The impact of jet energy scale uncertainties in event yields is evaluated by varying the jet energy corrections within their uncertainties and propagating the effect to the final result.
        The impact of the scale uncertainties on the category yields is largest for those targeting VBF, and can be as high as 15\%, but is less than 3\% for the resolution.
        The effect of the scale uncertainties is observed in Fig.~\ref{fig:VBF_score_and_valid} (right), where the combined uncertainty has large contributions from the jet energy scale uncertainty.
  \item \textit{Trigger efficiency}:
        the uncertainty in the efficiency of the trigger selection is measured with \Zee events using the T\&P technique.
        The size of its uncertainty is less than 1\%.
        An additional uncertainty is introduced to account for a gradual shift in the timing of the inputs of the ECAL L1 trigger in the region at $\abs{\eta} > 2.0$, which caused a specific trigger inefficiency during 2016--2017 data taking~\cite{CMS_L1T}.
\end{itemize}

The total impact on the \BHee measurement from experimental systematic uncertainties is $\pm{0.2\times10^{-4}}$, with the largest contributions arising from uncertainties affecting the electron energy scale.

\section{Results}
\label{sec:results}

This section describes the construction of the statistical model used in the maximum likelihood fit, and presents the results of the analysis.

\subsection{Signal models}

\begin{figure*}[htb!]
  \centering
  \includegraphics[width=0.49\textwidth]{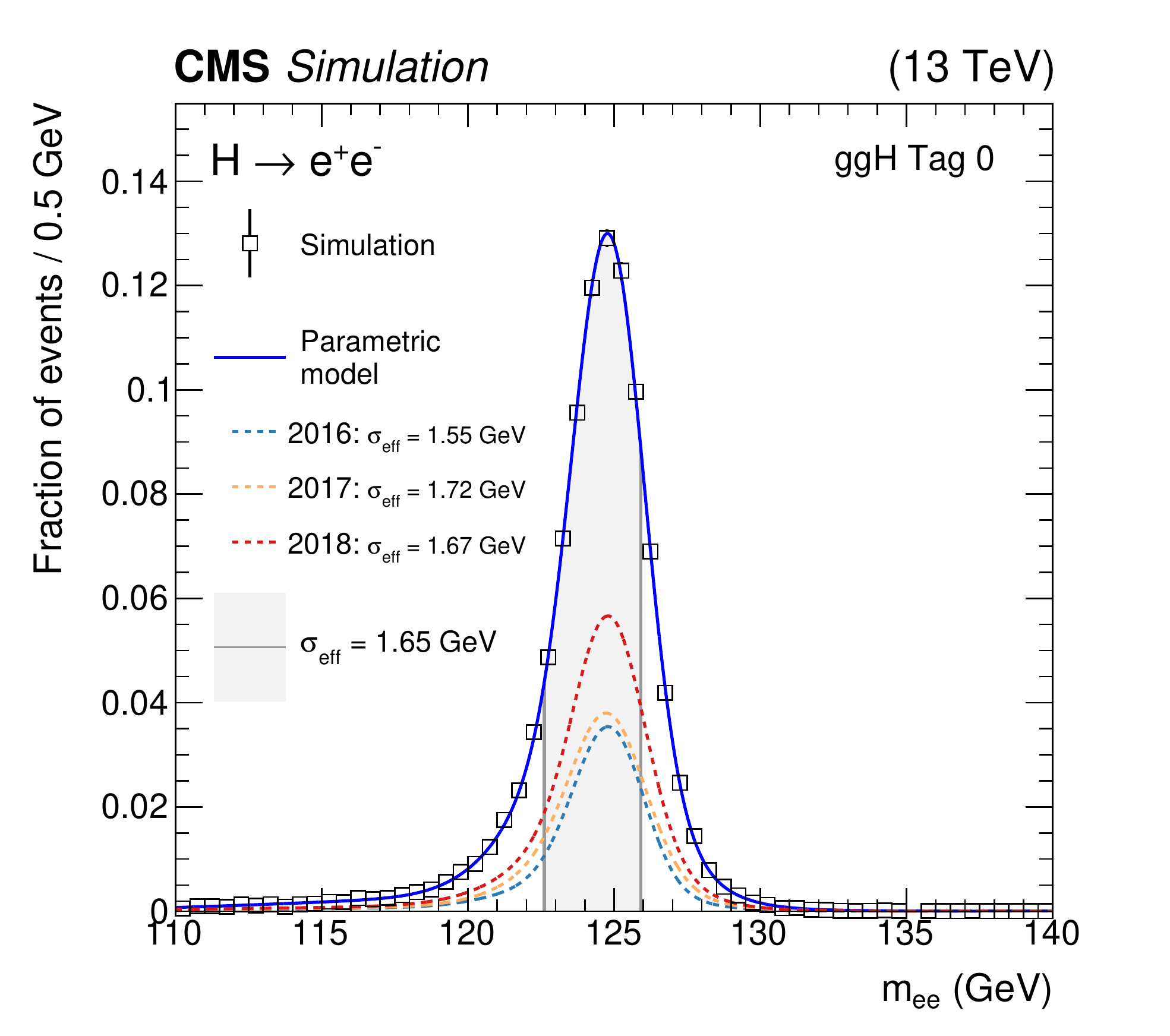}\hfill%
  \includegraphics[width=0.49\textwidth]{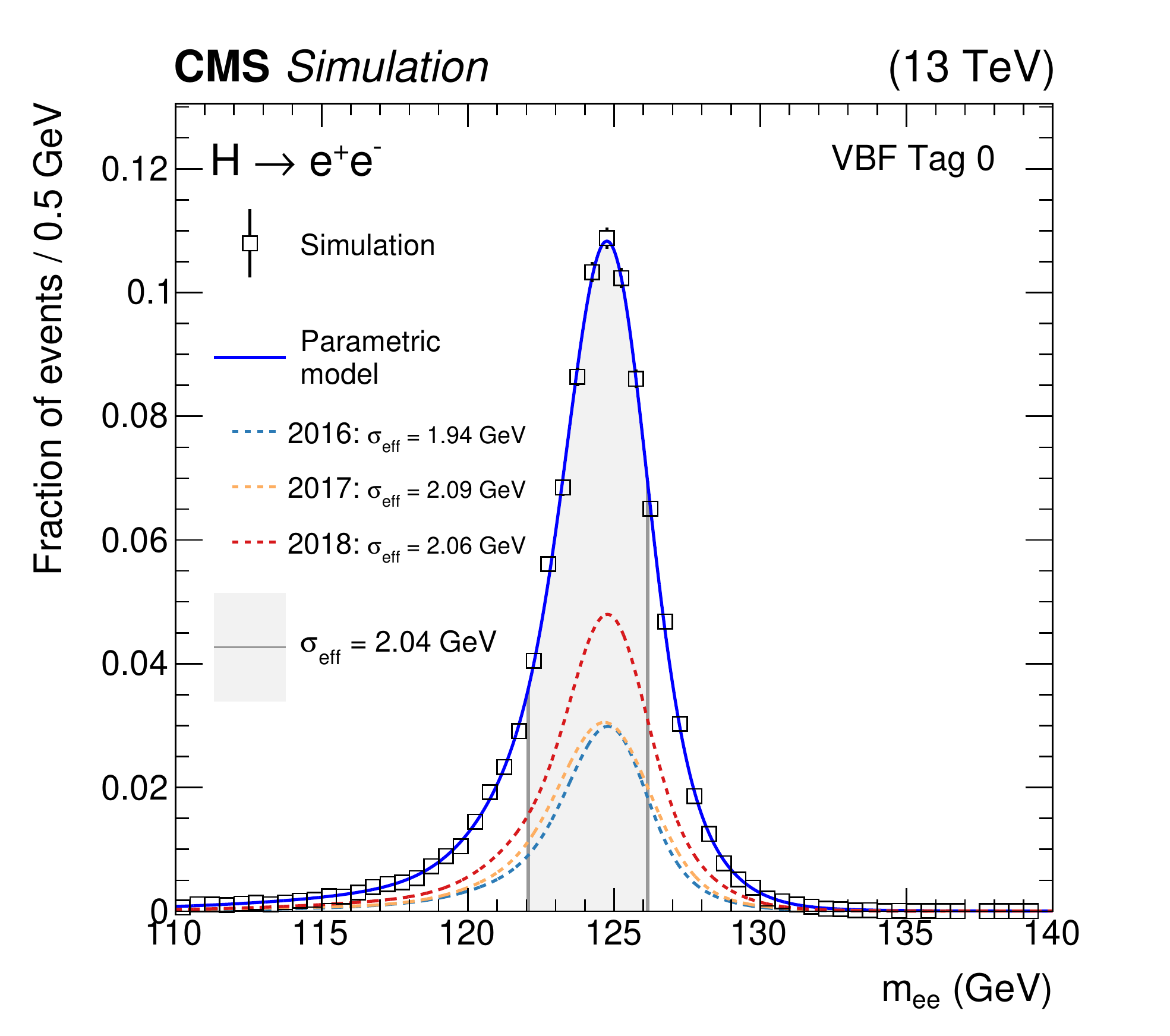}\hfill%
  \caption{
    Signal models for the highest S/B categories targeting \ggH and VBF processes, integrated over production processes, for Higgs boson events simulated at $\mH=125\GeV$.
    Contributions from each of the three years are shown by the dashed lines.
    The models are normalized to unit area.
    The \seff is the smallest interval containing 68.3\% of the \mee signal distribution.
  }
  \label{fig:smodels}
\end{figure*}

The shape of the \mee distribution for simulated signal is parameterized independently for each production process entering each event category, and for each year separately.
Each signal model is constructed from a sum of up to five independent Gaussian functions.
To allow for the extraction of limits on \BHee as a function of \mH, signal models are constructed from a simultaneous fit to samples at $\mH=120$, 125, and 130\GeV.
Each parameter for each Gaussian is a linear function of \mH.
The optimal number of Gaussian functions is chosen by performing an F-test~\cite{fTest}, to avoid overfitting to statistical fluctuations resulting from the limited size of the simulated samples.
The sum of models for each production process is shown for the highest S/B analysis categories targeting VBF and \ggH production in Fig.~\ref{fig:smodels}, for a nominal \mH of 125\GeV.

\subsection{Background models}

The discrete profiling method~\cite{envelope} is used to account for the uncertainty related to mismodelling of the background.
This method treats the choice of the function used to model the background, which is unknown a~priori, as a discrete nuisance parameter.
For each category, a range of analytical functions are used as candidates to fit the \mee distribution.
These include generic smoothly-falling functions such as sums of exponential functions, Laurent series', and Bernstein polynomials, as well as physics-inspired models chosen to fit the dominant DY component of the background, which comprise modified Breit--Wigner functions.
These functions have been used successfully to fit similar background processes in previous CMS analyses~\cite{HIG_19_015,CMSHMuMu}.
The final set of functions considered is chosen by performing an F-test for each candidate, with a loose requirement on the goodness-of-fit.

Each candidate function is considered in the final signal-plus-background (S+B) fit to data.
The function that results in the best overall fit is chosen for each value of the parameter of interest.
In the fit, the normalization and shape parameters for the background functions are allowed to vary freely.
The resulting confidence intervals therefore account for possible changes in the preferred function choice.
The bias on \BHee resulting from a particular choice of background function was also checked for each category and found to be negligible.

Finally, background events resulting from both SM \Hgg production, where the photons are misreconstructed as electrons, and from the Dalitz decay~\cite{Dalitz} (\Heeg), where the photon is included in the electron reconstruction, were studied with simulated events.
Each process was found to contribute around 0.1\% and 0.2\% to the inclusive analysis categories, respectively, for a \BHee at the expected limit.
These processes are, therefore, neglected in this analysis.

\subsection{Limits on \texorpdfstring{\BHee}{B(H to e+e-)}}

\begin{figure*}[htbp!]
  \centering
  \includegraphics[width=0.49\textwidth]{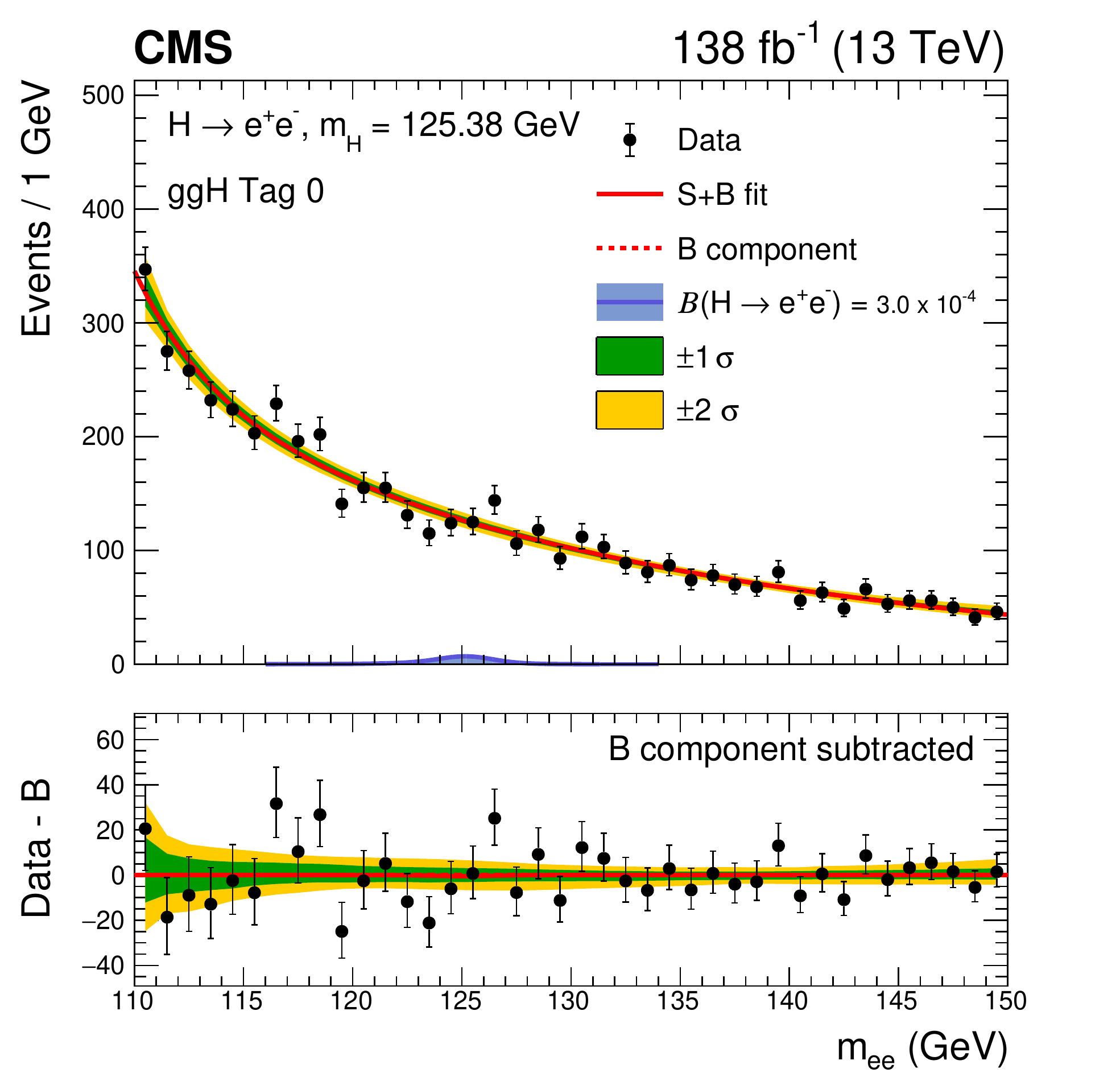}\hfill%
  \includegraphics[width=0.49\textwidth]{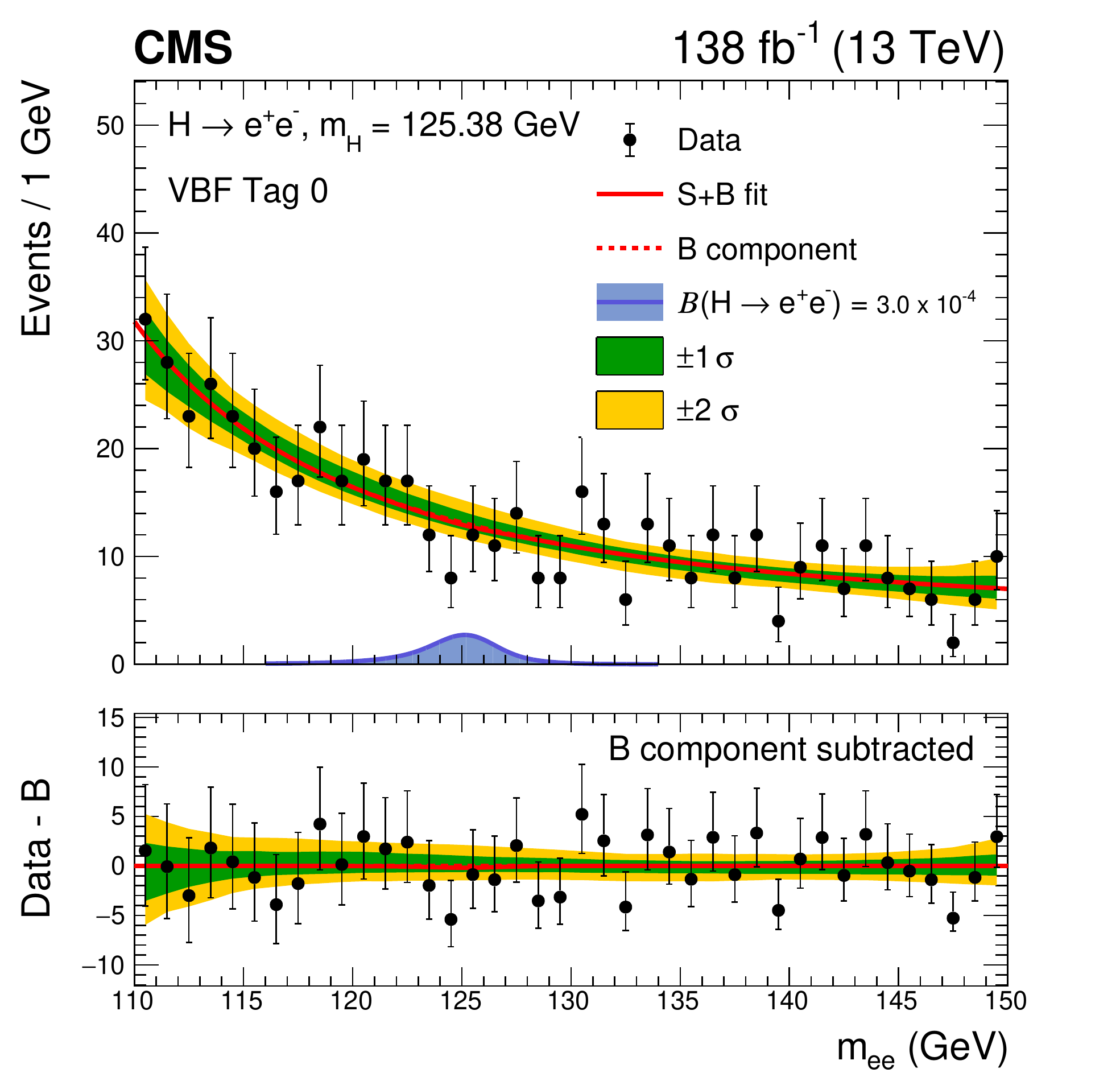}\hfill%
  \caption{
    The signal-plus-background model fit to the \mee distribution for the highest S/B analysis categories targeting the \ggH (left) and VBF (right) processes.
    The signal model for each category is also shown, scaled to the observed limit at $\mH=125.38\GeV$.
    The one (green) and two (yellow) standard deviation bands show the uncertainties in the background component of the fit.
    The lower panel shows the residuals after subtraction of this background component.
    The background functions describe the data well, with no excess observed.
  }
  \label{fig:S_plus_B_plots}
\end{figure*}

\begin{figure}[h!tb]
  \centering
  \includegraphics[width=\cmsFigWidth]{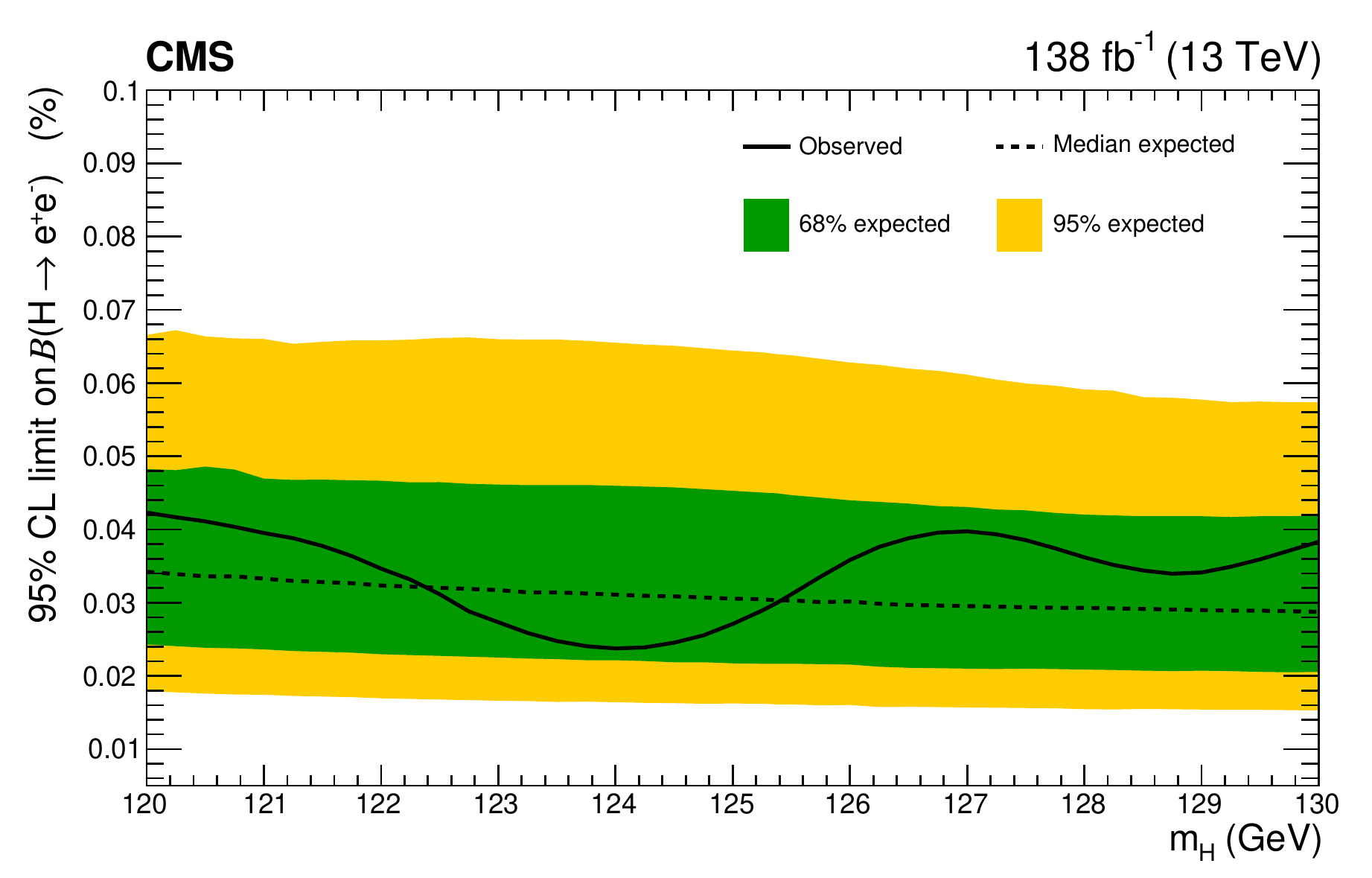}\hfill%
  \caption{Expected and observed limits on \BHee for a Higgs boson mass between 120--130\GeV.}
  \label{fig:limit_per_mass}
\end{figure}

\begin{figure}[h!tbp]
  \centering
  \includegraphics[width=\cmsFigWidth]{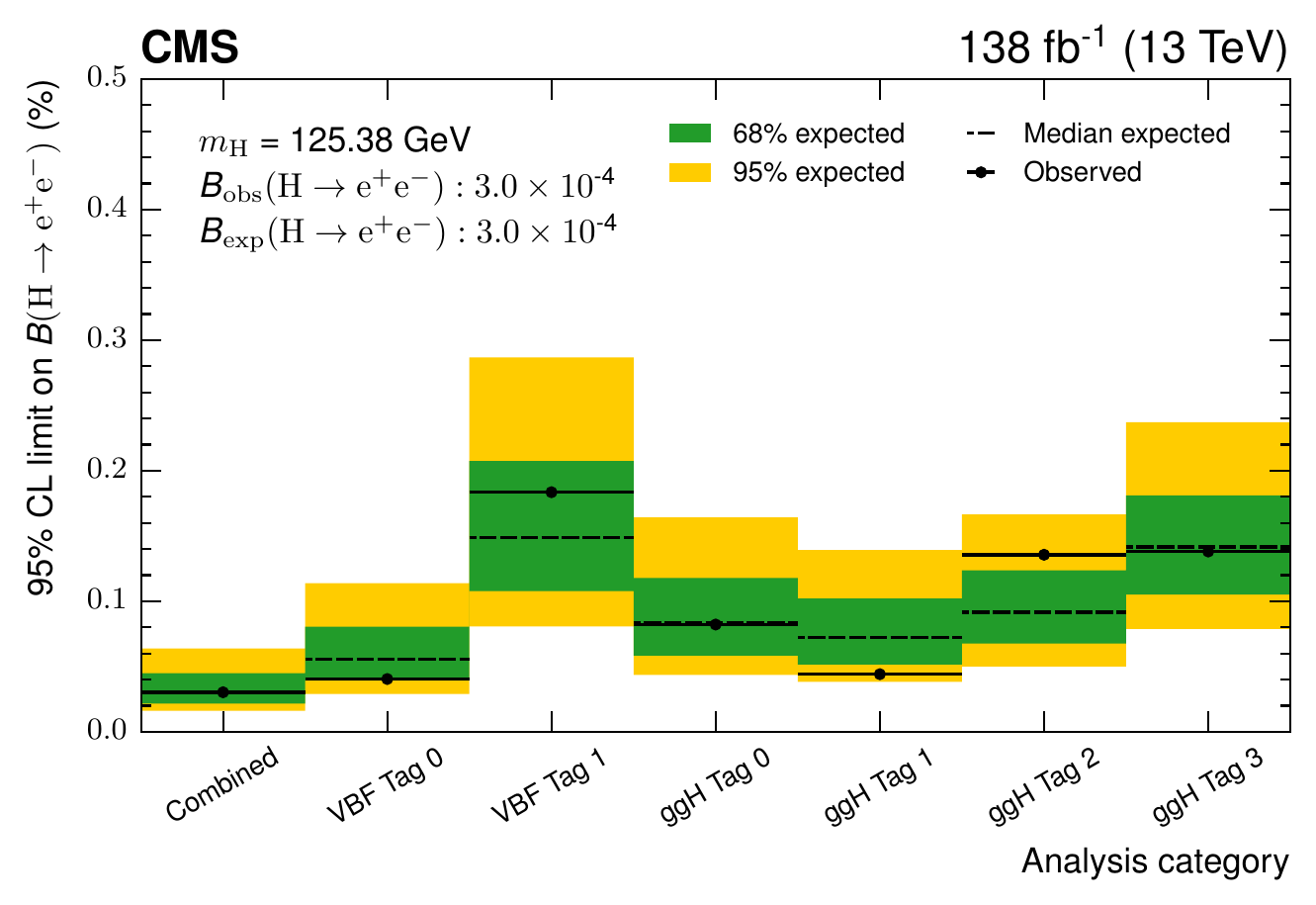}\hfill%
  \caption{
    Expected and observed limits on \BHee for each analysis category, and all categories combined
    The results here are computed for $\mH=125.38\GeV$.
  }
  \label{fig:limit_per_cat}
\end{figure}

A simultaneous maximum likelihood fit is performed across all analysis categories.
The (S+B) \mee distributions are shown for the highest S/B analysis categories targeting \ggH and VBF production in Fig.~\ref{fig:S_plus_B_plots}.
Upper limits are set on \BHee using the \CLs modified frequentist criterion~\cite{CLs1,CLs2,LHCStatProcedures} for \mH hypotheses between 120--130\GeV.
This construction uses the profile-likelihood ratio as the test statistic~\cite{Cowan}, under the asymptotic approximation.
Limits for a range of \mH hypotheses, along with the associated uncertainty intervals, are shown in Fig.~\ref{fig:limit_per_mass}.
At the current best measured value of $\mH=125.38\GeV$~\cite{CMS_Hgg_Hmass}, the observed (expected) 95\% CL limit is $\BHee < 3.0\times 10^{-4}$ ($3.0\times 10^{-4}$).
Assuming that the only deviation from the SM prediction is from a modified Yukawa coupling to the electron, the limit translates into an upper bound on the Higgs boson effective coupling modifier to electrons of $\abs{\kappa_{\Pe}}<240$.
A breakdown of the expected and observed limits on \BHee is shown per analysis category in Fig.~\ref{fig:limit_per_cat}.
Tabulated results are provided in the HEPData record for this analysis~\cite{hepdata}.

\section{Summary}
\label{sec:summary}

A search for the Higgs boson decaying to an \ee pair is performed using proton-proton collision data collected at \sqrts with the CMS experiment at the LHC between 2016--2018, corresponding to an integrated luminosity of \intlumi.
The analysis uses categories targeting Higgs boson production via gluon fusion and vector boson fusion, with dedicated boosted decision tree classifiers trained for each production mode to enhance the sensitivity of the resulting categories.
A maximum likelihood fit to the dielectron mass distribution is performed simultaneously in each analysis category to extract an upper limit on the Higgs boson to electron pair branching fraction; the resulting observed (expected) limit at the 95\% confidence level on the branching fraction for \Hee decays is $3.0\times 10^{-4}$ ($3.0\times 10^{-4}$).
This is the most stringent limit on the Higgs boson branching fraction to an \ee pair to date.
When compared with the previous best limit from the CMS Collaboration~\cite{CMSHeeRun1}, where analysis categories were constructed using a selection on electron and jet kinematics, the improvement in sensitivity presented in this Letter is attributed primarily to the use of BDT classifiers, which significantly improve the S/B ratio of analysis categories.
Accounting both for the increase in integrated luminosity and centre-of-mass energy, the use of BDT classifiers further improves the limit on \BHee by a factor of approximately 1.5.

\begin{acknowledgments}
We congratulate our colleagues in the CERN accelerator departments for the excellent performance of the LHC and thank the technical and administrative staffs at CERN and at other CMS institutes for their contributions to the success of the CMS effort. In addition, we gratefully acknowledge the computing centres and personnel of the Worldwide LHC Computing Grid and other centres for delivering so effectively the computing infrastructure essential to our analyses. Finally, we acknowledge the enduring support for the construction and operation of the LHC, the CMS detector, and the supporting computing infrastructure provided by the following funding agencies: BMBWF and FWF (Austria); FNRS and FWO (Belgium); CNPq, CAPES, FAPERJ, FAPERGS, and FAPESP (Brazil); MES and BNSF (Bulgaria); CERN; CAS, MoST, and NSFC (China); MINCIENCIAS (Colombia); MSES and CSF (Croatia); RIF (Cyprus); SENESCYT (Ecuador); MoER, ERC PUT and ERDF (Estonia); Academy of Finland, MEC, and HIP (Finland); CEA and CNRS/IN2P3 (France); BMBF, DFG, and HGF (Germany); GSRI (Greece); NKFIH (Hungary); DAE and DST (India); IPM (Iran); SFI (Ireland); INFN (Italy); MSIP and NRF (Republic of Korea); MES (Latvia); LAS (Lithuania); MOE and UM (Malaysia); BUAP, CINVESTAV, CONACYT, LNS, SEP, and UASLP-FAI (Mexico); MOS (Montenegro); MBIE (New Zealand); PAEC (Pakistan); MES and NSC (Poland); FCT (Portugal); MESTD (Serbia); MCIN/AEI and PCTI (Spain); MOSTR (Sri Lanka); Swiss Funding Agencies (Switzerland); MST (Taipei); MHESI and NSTDA (Thailand); TUBITAK and TENMAK (Turkey); NASU (Ukraine); STFC (United Kingdom); DOE and NSF (USA).
  
\hyphenation{Rachada-pisek} Individuals have received support from the Marie-Curie programme and the European Research Council and Horizon 2020 Grant, contract Nos.\ 675440, 724704, 752730, 758316, 765710, 824093, 884104, and COST Action CA16108 (European Union); the Leventis Foundation; the Alfred P.\ Sloan Foundation; the Alexander von Humboldt Foundation; the Belgian Federal Science Policy Office; the Fonds pour la Formation \`a la Recherche dans l'Industrie et dans l'Agriculture (FRIA-Belgium); the Agentschap voor Innovatie door Wetenschap en Technologie (IWT-Belgium); the F.R.S.-FNRS and FWO (Belgium) under the ``Excellence of Science -- EOS" -- be.h project n.\ 30820817; the Beijing Municipal Science \& Technology Commission, No. Z191100007219010; the Ministry of Education, Youth and Sports (MEYS) of the Czech Republic; the Hellenic Foundation for Research and Innovation (HFRI), Project Number 2288 (Greece); the Deutsche Forschungsgemeinschaft (DFG), under Germany's Excellence Strategy -- EXC 2121 ``Quantum Universe" -- 390833306, and under project number 400140256 - GRK2497; the Hungarian Academy of Sciences, the New National Excellence Program - \'UNKP, the NKFIH research grants K 124845, K 124850, K 128713, K 128786, K 129058, K 131991, K 133046, K 138136, K 143460, K 143477, 2020-2.2.1-ED-2021-00181, and TKP2021-NKTA-64 (Hungary); the Council of Science and Industrial Research, India; the Latvian Council of Science; the Ministry of Education and Science, project no. 2022/WK/14, and the National Science Center, contracts Opus 2021/41/B/ST2/01369 and 2021/43/B/ST2/01552 (Poland); the Funda\c{c}\~ao para a Ci\^encia e a Tecnologia, grant CEECIND/01334/2018 (Portugal); the National Priorities Research Program by Qatar National Research Fund; MCIN/AEI/10.13039/501100011033, ERDF ``a way of making Europe", and the Programa Estatal de Fomento de la Investigaci{\'o}n Cient{\'i}fica y T{\'e}cnica de Excelencia Mar\'{\i}a de Maeztu, grant MDM-2017-0765 and Programa Severo Ochoa del Principado de Asturias (Spain); the Chulalongkorn Academic into Its 2nd Century Project Advancement Project, and the National Science, Research and Innovation Fund via the Program Management Unit for Human Resources \& Institutional Development, Research and Innovation, grant B05F650021 (Thailand); the Kavli Foundation; the Nvidia Corporation; the SuperMicro Corporation; the Welch Foundation, contract C-1845; and the Weston Havens Foundation (USA).
\end{acknowledgments}

\bibliography{auto_generated}
\cleardoublepage \appendix\section{The CMS Collaboration \label{app:collab}}\begin{sloppypar}\hyphenpenalty=5000\widowpenalty=500\clubpenalty=5000
\cmsinstitute{Yerevan Physics Institute, Yerevan, Armenia}
{\tolerance=6000
A.~Tumasyan\cmsAuthorMark{1}\cmsorcid{0009-0000-0684-6742}
\par}
\cmsinstitute{Institut f\"{u}r Hochenergiephysik, Vienna, Austria}
{\tolerance=6000
W.~Adam\cmsorcid{0000-0001-9099-4341}, J.W.~Andrejkovic, T.~Bergauer\cmsorcid{0000-0002-5786-0293}, S.~Chatterjee\cmsorcid{0000-0003-2660-0349}, K.~Damanakis\cmsorcid{0000-0001-5389-2872}, M.~Dragicevic\cmsorcid{0000-0003-1967-6783}, A.~Escalante~Del~Valle\cmsorcid{0000-0002-9702-6359}, P.S.~Hussain\cmsorcid{0000-0002-4825-5278}, M.~Jeitler\cmsAuthorMark{2}\cmsorcid{0000-0002-5141-9560}, N.~Krammer\cmsorcid{0000-0002-0548-0985}, L.~Lechner\cmsorcid{0000-0002-3065-1141}, D.~Liko\cmsorcid{0000-0002-3380-473X}, I.~Mikulec\cmsorcid{0000-0003-0385-2746}, P.~Paulitsch, F.M.~Pitters, J.~Schieck\cmsAuthorMark{2}\cmsorcid{0000-0002-1058-8093}, R.~Sch\"{o}fbeck\cmsorcid{0000-0002-2332-8784}, D.~Schwarz\cmsorcid{0000-0002-3821-7331}, S.~Templ\cmsorcid{0000-0003-3137-5692}, W.~Waltenberger\cmsorcid{0000-0002-6215-7228}, C.-E.~Wulz\cmsAuthorMark{2}\cmsorcid{0000-0001-9226-5812}
% ...
\par}
\vskip\cmsinstskip
\dag:~Deceased\\
$^{1}$Also at Yerevan State University, Yerevan, Armenia\\
$^{2}$Also at TU Wien, Vienna, Austria\\
% ...
\end{sloppypar}
%%% END EDITABLE REGION %%%
% skeleton_end
\end{document}